\documentstyle[preprint,aps,floats,psfig]{revtex}
\bibstyle{unsrt}

\newcommand{\ls}{a_s}
\newcommand{\lt}{a_t}

\tighten
\begin{document}
\draft
 
\preprint{MIT-CTP-3150}
 
\title{Langevin Evolution of Disoriented Chiral Condensate}
\author{Lu\'{\i}s~M.~A.~Bettencourt$^{1,2}$, Krishna~Rajagopal$^1$ 
and James~V.~Steele$^1$ }
\address{$^1$ Center for Theoretical Physics, Massachusetts Institute 
of Technology, Cambridge MA 02139}
\address{$^2$ Theoretical Division, MS B288, Los Alamos National Laboratory,
Los Alamos NM 87545}
\date{\today}
\maketitle

\begin{abstract}
As the matter produced in a relativistic heavy ion collision
cools through the QCD phase transition, the dynamical
evolution of the chiral condensate will be driven out
of thermal equilibrium.  As a prelude to analyzing this
evolution, and in particular as a prelude
to learning how rapid the cooling must be in order for
significant deviations from equilibrium to develop, 
we present a detailed analysis of the time-evolution
of an idealized region of disoriented chiral condensate.
We set up a Langevin field equation which can describe the
evolution of these (or more realistic) linear
sigma model configurations
in contact with a heat bath representing the presence of other
shorter wavelength degrees of freedom.
We first analyze the
model in equilibrium, paying particular attention
to subtracting ultraviolet divergent classical terms
and replacing them by their finite quantum counterparts.
We use known results from lattice gauge theory and chiral
perturbation theory to fix nonuniversal constants.
The result is a theory which is ultraviolet cutoff independent 
and that reproduces quantitatively the expected equilibrium
behavior of the quantum field theory 
of pions and $\sigma$ fields over a wide range of 
temperatures.
Finally, we estimate the viscosity $\eta(T)$, which controls
the dynamical timescale in the Langevin equation,
by requiring that the timescale for DCC decay agrees with
previous calculations.  
The resulting $\eta(T)$ is larger
than that found perturbatively.
We also determine the temperature below which the classical field Langevin 
equation ceases to be a good model for the quantum field dynamics.
\end{abstract}


\section{Introduction}

The relativistic heavy-ion collider (RHIC) is now creating
strongly-interacting matter at high enough energy density
that there is every expectation that temperatures 
above the QCD phase transition are being explored.
The question of equilibration is
crucial to the understanding of these collisions.
There has been much progress on one aspect
of this question recently: we now know that (at least
at arbitrarily high collision energies) an equilibrated
quark-gluon plasma {\it will} be created~\cite{MuellerSon}.
Furthermore, the first analyses of elliptic flow data 
can be interpreted as indicating early thermalization in RHIC
collisions~ \cite{EllipticFlow}.
What remains quite unclear, however, is 
whether even if an equilibrated partonic medium 
is attained early in the collision, thermal equilibrium
is maintained as the matter cools through the QCD phase transition and
becomes a gas of hadrons, which subsequently freeze out
and are seen in a detector.  

The QCD phase transition is likely a smooth crossover
if it is traversed at low baryon number chemical potential $\mu_B$.
At larger $\mu_B$, it is thought to be first order.  
This means that  at a critical $\mu_B$,
there is a second order phase 
transition in the universality class
of the three-dimensional Ising 
model~\cite{earlycritptpapers,SRS1,SRS2,critptreview}.  
The nature of the possible nonequilibrium
phenomena are quite different in these three cases~\cite{Bubbling}.
If the transition is first order, bubbling may yield a spatially
inhomogeneous final state~\cite{HeiselbergJacksonMishustin}.  
If the transition occurs near
the second order critical point, then nonequilibrium
effects tend to obscure the unique fluctuations
characteristic of the equilibrium critical point:  in
equilibrium, the correlation length would diverge there;
nonequilibrium effects smooth out this divergence~\cite{BerdnikovRajagopal}.
The farther from equilibrium the evolution is, the
less likely are distinctive signatures of the critical point.
Finally, if the transition is a crossover, as is likely
at RHIC where $\mu_B$ is small, and if this crossover
is traversed sufficiently quickly that the dynamics
can be treated as a ``maximally out of equilibrium'' quench,
then long wavelength oscillations of the chiral condensate
are excited to amplitudes which are greatly in excess
of those present in equilibrium~\cite{RajagopalWilczekDCC}.  
The primary signal
of these disorientations of the chiral condensate (DCC)
is a large number of soft pions exhibiting large fluctuations in 
isospin~\cite{AnselmRyskin,BlaizotKrzywicki,Bjorken,RajagopalWilczekStatic,RajagopalWilczekDCC},
which have been looked for and not seen in lower energy heavy
ion collisions at the CERN SPS~\cite{WA98}.
Many other DCC signatures have also been discussed~\cite{OtherSignatures}. 
The crucial unresolved
question in all three cases is: ``How rapid must the cooling
be in order for the evolution to deviate significantly from
thermal equilibrium?''  
In this paper, we lay the
ground work for answering this question in the case
of a smooth crossover. 

We construct a Langevin description of the nonequilibrium
dynamics of the chiral condensate which we can use over
a range of temperatures around the phase transition.
Our dynamical degrees of freedom are those of the
classical linear sigma model.  This is valid as long
as $m_\sigma$ and $m_\pi$ are much smaller than the temperature.
It is therefore valid near $T_c$ sufficiently close to the chiral limit.
We shall use it with parameters chosen to reproduce
the physical pion mass at zero temperature, and this will mean
that we are using the classical theory beyond its realm of
quantitative validity.  Until quantum field
theoretical approaches (for 
example those of Refs.~\cite{NonEqbmQFT}) mature
to the point where they can be used to describe 
spatially varying field configurations, the present
classical analysis is of value as a guide.  

Why introduce a Langevin heat bath? That is, why go beyond
the purely classical analysis of Refs.~\cite{RajagopalWilczekDCC,AHW}?  
Allowing
the long wavelength degrees of freedom to exchange energy
with a heat bath is a crude way of representing the
existence of other degrees of freedom.  Our reason
for introducing it is that we wish to investigate
the response of the long wavelength degrees of freedom to decreases in
the temperature $T$ of the heat bath that occur with
varying speeds. If $T$ is reduced arbitrarily slowly,
the system must stay arbitrarily close to equilibrium.
If $T$ is reduced suddenly, nonequilibrium dynamics
results.  With a Langevin equation, therefore, we
shall be able to learn how fast the cooling must
be in a heavy ion collision if the traversal of
the crossover region of the phase transition is to
be associated with DCC phenomena.  In addition to a 
time varying heat bath temperature, we shall also be able
to use the freedom to let $T$ vary as a function of position
to consider initial conditions in which the hot region
is finite in extent. 

We leave 
the answer to the questions just posed to a subsequent paper.
Here, we focus on setting up the Langevin description,
and choosing all the associated parameters.  
Many parameters can be chosen entirely with reference
to properties of the system in thermal equilibrium.
We can fix these parameters by comparison with known results 
from chiral perturbation theory and lattice gauge theory.
In so doing we obtain a Langevin equation which gives a 
reasonable description of the equilibrium physics
from very low temperatures up to temperatures well
above $T_c$.  This is possible in equilibrium, but 
the classical description of {\it dynamics} must break 
down at low temperatures, where $m_{\sigma,\pi}\gg T$.
We discover this breakdown when we seek to fix those parameters
in the Langevin equation which require dynamical input,
namely the viscosity $\eta(T)$. To this
end, we analyze the decay of an idealized DCC
with time, due to the presence of the heat bath.
This problem has been analyzed previously 
by Steele and Koch~\cite{SteeleKoch},
and we use their results to fix the viscosity in
our Langevin equation.  
We find that in the vicinity of $T_c$, where we 
expect our treatment to be valid, we can successfully
reproduce the DCC decay timescale of Ref.~\cite{SteeleKoch}.
In order to do so, the $\eta$ we introduce
is about a factor of five larger than that predicted
by perturbative calculation~\cite{Rischke:1998qy}. 
At lower temperatures, the whole analysis breaks down:
even with the dissipation
due to the heat bath completely removed by setting
$\eta$ to zero,
we find that the DCC decays much more quickly than 
it should. The long wavelength DCC is decaying not
via interaction with the  stochastic heat bath but via
its interactions with 
modes which we are treating classically.
We are therefore able to evaluate the temperature
below which our classical Langevin field equation should not
be used to model the dynamics. 

Our analysis complements that in several
previous papers. Greiner and M\"uller~\cite{Greiner:1997dx} 
and Rischke~\cite{Rischke:1998qy} have derived effective Langevin 
equations by integrating out hard modes perturbatively and found 
that these typically include correlated and multiplicative sources 
of noise and dissipation.
Xu and  Greiner~\cite{XuGreiner} then 
explored the effects of such
noise and dissipation on $0+1$-dimensional dynamics.
This treatment of the noise goes beyond ours, since 
we treat the heat bath as an uncorrelated source of fluctuation.  
However, whereas $0+1$-dimensional calculations (like those
of Refs.~\cite{XuGreiner,BiroGreiner})
treat the dynamics of only one
mode explicitly and thus assume spatial homogeneity, 
our $3+1$-dimensional Langevin equation
allows an analysis of more realistic, spatially varying,
configurations.  The $3+1$-dimensional Langevin equation
has been introduced previously by Chaudhuri~\cite{Chaudhuri1},
although as he notes in Ref.~\cite{Chaudhuri2} all of his
results depend sensitively on his choice of lattice spacing.
We introduce the correct $T$-dependent counterterms in order
to obtain results at long wavelengths 
which are independent of the lattice-spacing.
Randrup has addressed the linear sigma model dynamics 
in a series of papers, using either the classical theory or 
a Hartree approximation~\cite{Randrup}. 
These studies have shown the importance
of accounting correctly for the multidimensional expansion of the hot 
system, and have provided initial answers
to some of the questions we wish to investigate. The
stochastic linear sigma model we develop below has
the relative advantage of having dynamics which equilibrate
at late times in the absence of expansion and in addition
allows us to take some account of the effects of
shorter wavelength degrees of freedom on the longer
wavelength modes we treat classically. 

In Section II we describe the model, paying close attention
to the counterterms which are needed in order to ensure
that the long wavelength physics is not divergent as
the lattice spacing is taken to zero.  In Section III,
we evaluate the order parameter as a function of temperature
and fix the most important finite counterterms by enforcing agreement
with chiral perturbation theory at low temperature and with
lattice gauge theory calculations of the critical temperature.
We also evaluate the temperature-dependent pion and sigma masses.
In Section IV, we analyze the decay of a DCC. By comparing
the DCC lifetime we obtain to that calculated previously
by other methods, we fix the viscosity $\eta$ in the Langevin
equation, which couples the classical fields to the heat bath.
We also determine the temperature below which a classical field
treatment fails to describe the dynamics correctly.
We close in Section V with a look towards the future.

\section{Langevin Equations and Counterterms}

At low temperatures, the correct effective field theory for
QCD is the nonlinear sigma model, describing the dynamics of
the pseudoscalar pions.  At the QCD phase transition, however, chiral
symmetry is approximately restored and the scalar sigma 
becomes approximately degenerate with the pions.  These
four degrees of freedom are the lightest 
in QCD near its finite temperature phase transition.
Indeed, if the up and down quarks were 
massless, these four modes would all be massless at $T_c$,
making the transition an $O(4)$ second order 
phase transition~\cite{PisarskiWilczek,RajagopalWilczekStatic}.
Lattice QCD
calculations suggest that with quark masses as in nature, 
these modes have inverse correlation lengths which are 
comparable to $T_c$~\cite{LatticeScreeningLengths}.
This means that although these modes are still the lightest
degrees of freedom, the classical treatment we employ
is at or beyond the limit of  its regime of validity.
Two asides should be noted at this point. First, at
the $\mu_B$ at which there is a second order critical
point in the QCD phase transition, the sigma becomes
massless while the pions do not.  Second,  there have
been recent suggestions that the lightest scalar
glueball may play an interesting role at the transition~\cite{Pisarski}.
Lattice QCD calculations show that this has a 
correlation length close to $(3 T_c)^{-1}$~\cite{Gupta}, much
shorter than that of the pion~\cite{GavaiGupta},
and we therefore neglect 
it.

Writing the pions and the sigma as an $O(4)$ vector
$\phi_a=(\sigma,{\vec \pi})$, the linear sigma model Lagrangian
is given by
\begin{equation}
{\cal L} = \frac12 \partial_\mu \phi_a \partial^\mu \phi_a -
\frac{\lambda}4 \left( \phi_a^2 - v^2 \right)^2 + H \sigma,
\label{lagrangian}
\end{equation}
with summation over repeated indices implied.
The context in which this is of precise validity is 
two-flavor QCD in the chiral limit, on the assumption
that this theory has a second order phase transition,
as is consistent with what we know from lattice QCD
calculations~\cite{LatticeReviews}.
Then, that transition
is in the same $O(4)$ universality class as the second
order phase transition found at nonzero temperature 
in the theory (\ref{lagrangian}) with 
$H=0$~\cite{PisarskiWilczek,RajagopalWilczekStatic}.
The explicit symmetry breaking term $H$, the coupling $\lambda$, and
the vacuum expectation value $v$ can be expressed
in terms of the phenomenological parameters $m_\sigma$, $m_\pi$, and
$f_\pi$ at tree level via
\begin{equation}
\!\!H=f_\pi m_\pi^2 \ ,\quad\!
\lambda = \frac{m_\sigma^2 - m_\pi^2}{2 f_\pi^2} \ , \quad\!
v^2 = \frac{m_\sigma^2 - 3 m_\pi^2}{m_\sigma^2 - m_\pi^2} f_\pi^2  \ .
\label{params}
\end{equation}
We take $(f_\pi,m_\pi,m_\sigma)=(92.4,135,600)$~MeV,
meaning $H=(119. {\rm ~MeV})^3$, $\lambda=20.0$
and $v=87.3 {\rm ~MeV}$.

The simplest Langevin prescription corresponds to including additive 
stochastic 
sources $\xi_a(x,t)$ and dissipation terms in the Euler-Lagrange 
equations of motion for the fields:
\begin{equation}
\frac{\partial^2\phi_a}{\partial t^2} - \nabla^2 \phi_a
+ \lambda \left( \phi_a^2 - v^2 \right) \phi_a - H \delta_{a0}
= -\eta_{ab} \frac{\partial \phi_b}{\partial t}  + \xi_a \ .
\label{langevin}
\end{equation} 
The stochastic fields $\xi_a$ are taken to obey the fluctuation-dissipation 
relations
\begin{equation}\label{fluctuationdissipation}
\langle \xi_a(x) \xi_b(y) \rangle = 2 \eta_{ab}(x) T 
\;\delta^4(x-y) \ ,
\end{equation}
with $\langle \xi_a(x) \rangle =0$. In this paper, we shall always
choose $\eta_{ab}(x)=\eta(T)\delta_{ab}$.
This dynamics reduces to the classical microcanonical relativistic 
evolution for the pions and sigma in the limit of small 
dissipation $\eta\rightarrow 0$.
In the limit of vanishing masses the long wavelength modes of the fields
are effectively overdamped, and suffer critical slowing down~\cite{Bett}. 
Then our equations correspond to Model A, in the classification of 
dynamical stochastic theories of Hohenberg and Halperin~\cite{HoHa}. 
Although Model A will be adequate for our purposes, a complete analysis
of the universal $O(4)$ dynamics in the chiral limit of QCD
requires coupling the order parameter (which is not
conserved) to other 
conserved quantities in the theory, as in Hohenberg and Halperin's 
Model G~\cite{RajagopalWilczekStatic}.

For any nonzero
value of  $\eta$, at late times ($t\gg\eta^{-1}$)
the Langevin dynamics describes evolution
towards a steady state 
characterized by the equilibrium 
partition function
\begin{eqnarray}
{\cal Z} = N \int D \phi E^{-\beta {\cal H}[\phi]},
\label{partfunction}
\end{eqnarray}
with 
\begin{eqnarray}
{\cal H}[\phi] = \int d^3 x~ \left[ \frac12 \left(\nabla \phi_a \right)^2
+ \frac{\lambda}4 \left( \phi_a^2 - v^2 \right)^2 - H \sigma \right].
\end{eqnarray}
Eq.~(\ref{partfunction}) is the canonical partition function for 
the linear sigma model in 3D. Note that the ensemble of equilibrium
configurations described by this partition function is independent
of $\eta$.  In this section and the next, we shall use equilibrium
physics to fix all parameters in the model except $\eta$. In
Section IV, we shall fix $\eta$ by analyzing the dynamics
of the relaxation of nonequilibrium configurations towards 
equilibrium.

Eq.~(\ref{partfunction}) is the correct equilibrium state 
for the {\it classical} model. When compared to the quantum partition function
it fails to account for the correct statistical 
weighting of hard particles (with typical momentum $k\sim T$) and 
at low temperatures, when ($T < m_\pi,m_\sigma$). We will return 
to the latter point in the next section.
The classical statistical weighting of hard modes in the equilibrium state 
leads to ultraviolet divergences (as the lattice spacing $a_s$ is taken 
to zero) of certain expectation values at nonzero temperature. 
This effect is an incorrect prediction of the classical 
partition function.  Fortunately it can be compensated for. The idea 
is to subtract the divergent terms, which are all perturbative and 
can therefore be identified diagrammatically.  These are then replaced by the 
corresponding
correct contributions, identified through the computation of the same 
diagram using quantum statistical distributions.

\begin{figure}
\begin{center}
\leavevmode
\psfig{file=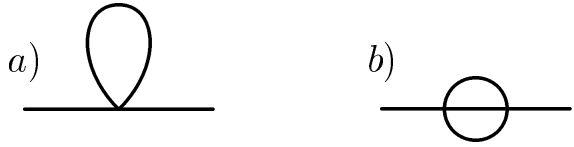,width=4.in,silent=}
\end{center}
\caption{The two perturbative diagrams that are ultraviolet divergent
in the classical theory at finite temperature: $a)$ the tadpole and 
$b)$ the sunset. Both divergences are momentum independent allowing us 
to compute the sunset diagram at zero external momentum only.}
\label{figdiag}
\end{figure}
For the linear sigma model in three
dimensions, there are only two divergent diagrams, shown in 
Fig.~\ref{figdiag}. These are self-energy contributions:
the tadpole diagram at one loop and the sunset diagram at two
loops~\cite{Parisi}. Both divergences are momentum independent and 
can be subtracted away by a simple temperature dependent mass renormalization. 
These two diagrams can be written as
\begin{eqnarray}
I_{\rm tadpole} &=& \frac{4!}{2!}\; \frac{\lambda}4 \; \frac{N+2}3
\; \int \!\! {d^3k \over (2 \pi)^3} \; G(\omega_k ),
\\
I_{\rm sunset} &=& \frac{4!^2}{3!} \left(\frac{\lambda}4 \right)^{\!\!2}
\frac{N+2}3  
\int \!\! {d^3k \, d^3q \over (2 \pi)^6} 
\frac{G(\omega_k) G(\omega_q)}
{\vert \vec k +\vec q \vert^2 + m^2} \ .
\end{eqnarray}
We have made explicit all symmetry factors.
The two-point function for the scalar field
$G(\omega)$ has the thermal (no vacuum contributions) form 
$G(\omega)=n(\omega)/\omega$, with the single particle number
distribution given by 
\begin{eqnarray}
n(\omega) = \left \{
\begin{array}{ll}
\displaystyle
\frac{T}{\omega}\ , & \mbox{classical}\ ,
\\[3 ex]
\displaystyle
\frac{\hbar}{e^{\hbar\omega/T}-1}\ , 
\qquad              &\mbox{Bose-Einstein}\ ,
\end{array}
\right.
\label{nw}
\end{eqnarray}
for classical or quantum scalar fields respectively.
Clearly, the quantum distribution approaches the classical for 
$\hbar\omega/T \rightarrow 0$, i.e., in the limit of high occupation 
numbers. Ultimately this is why the classical theory becomes a 
good description 
of long wavelength physics when masses are vanishing or at least small 
compared to $T$, as happens at a true critical point.

On the lattice, the continuum 
dispersion relation for 
a particle of mass $m$, namely $\omega^2=k^2+m^2$,
becomes
\begin{eqnarray}
\omega^2 = \frac{4}{\ls^2} \sum_{i=1}^{D} \sin^2 \!\!\left( \frac{k_i\,
\ls}{2} \right) + m^2 \ ,
\label{omegalattice}
\end{eqnarray}
where $\ls$ is the lattice spacing and  $D$ the number of spatial dimensions.
Here, $k_i=2\pi n_i/N_i \, \ls$, with $N_i$ the linear
size of the lattice and $n_i \in \{-\frac12 N_i,\frac12 N_i\}$, for each
Euclidean dimension. Eq.~(\ref{omegalattice})
reduces to the continuum result when $k_i\, \ls$ is small.

If we are interested only in the ultraviolet behavior of the 
tadpole diagram Fig.~1a, 
we can neglect the mass $m$. (This 
diagram has no infrared divergences.) 
This yields
\begin{eqnarray}
I_{\rm tadpole}^{\rm cl} &=& \frac{N+2}{2\pi^2} \lambda T \Lambda
\;\to\; 0.25\, \frac{(N+2)\lambda T}{\ls} \ ,
\\
I_{\rm tadpole}^{\rm B-E} &=& \frac{N+2}{12} {\lambda T^2 \over \hbar}
\ .
\end{eqnarray}
The classical tadpole is linearly divergent with the ultraviolet 
momentum cutoff 
$\Lambda$.  The arrow shows the result on the lattice, from
summing over modes within a Brillouin zone. 
To the accuracy quoted, these lattice results are identical for $N_i\ge32$ 
and various temperatures.
The Bose-Einstein tadpole is finite. 

The sunset diagram Fig.~\ref{figdiag}b, has only a logarithmic divergence 
and so in addition to being ultraviolet divergent it 
is infrared divergent in the absence of masses~\cite{Farakos}.
Since only the difference of the two contributions 
$I^{\rm B-E}-I^{\rm cl}\equiv \Delta I$ enters our calculation,
the infrared divergence cancels
out (allowing us again to neglect the mass $m$) and
the ultraviolet divergence is 
$(N+2) \lambda^2 T^2 \ln(T/\hbar \Lambda)/8 \pi^2$.
There is no convenient analytic form for the finite part, but on the 
lattice the result is 
\begin{eqnarray}
\Delta I_{\rm sunset} \to \left[0.0144\, \ln (T \ls/\hbar) - 0.0369\right]
  (N+2) \lambda^2 T^2  \ .
\end{eqnarray}

The renormalization is then achieved by making the replacement
\begin{eqnarray}
-\lambda v^2 \rightarrow 
-\lambda v^2 + \Delta I_{\rm tadpole} + \Delta I_{\rm sunset}
\label{counterterms}
\end{eqnarray}
in the term linear in $\phi_a$ in the equation
of motion Eq.~(\ref{langevin}),
which removes the linear and logarithmic divergences in the self energy.
Note that $\Delta I\to 0$ as $T\to 0$, and so
this renormalization does not affect the choice of 
the parameters in Eq.~(\ref{params}) at zero temperature.
This is true to all loop orders.
The procedure we have described
implements the matching of the classical thermal theory 
to the quantum in the ultraviolet, where the classical partition function 
explicitly leads to divergences.  After adding the
counterterms (\ref{counterterms}), the long wavelength
classical physics is non-divergent as the lattice spacing
is taken to zero.  Furthermore, these counterterms alone
ensure that at high temperature, the long wavelength physics
is correct and independent of the ultraviolet cutoff.

At low temperatures, however, differences between
the classical and quantum theories which do not
diverge with the ultraviolet cutoff 
become important.  Such ultraviolet-finite $T$-dependent counterterms arise
from many diagrams, not just from the two in Fig. 1.
We will deal with this pragmatically, by requiring that 
the behavior of our partition function matches the predictions of 
chiral perturbation theory for $T \ll f_\pi$ and shows a critical
temperature $T_c$ in agreement with that found in lattice
calculations of full QCD.  In
the next section, we shall use these two criteria
to fix the coefficients
of finite counterterms which are linear and quadratic
in $T$.  

For numerical work, it is convenient to remove the large coupling
$\lambda\sim 20$ from the Langevin evolution and express all quantities
in dimensionless units by scaling to new primed quantities:
\begin{eqnarray}
&& x' = \sqrt{\lambda}\; v\, x, \qquad t' = \sqrt{\lambda} \; v \; t, \\
&& \phi' ={\phi \over v}, \qquad \qquad
H'={H\over \lambda v^3} \\ 
&& \eta ' = {\eta \over \sqrt{\lambda} \; v}, \qquad \quad
T' = {\sqrt{\lambda} \over v} T.
\label{rescal}  
\end{eqnarray}
Using Eq.~(\ref{params}) gives $H'=0.1263$.
We must be careful with these rescalings since they are inconsistent
with the conventional choice $\hbar=1$.
The action scales as $S\to S'/\lambda$ and so 
in order to leave the path integral invariant,
Planck's constant
$\hbar$ must also be scaled:
\begin{equation}
\hbar' = \lambda \hbar \ .
\end{equation}
The tadpole and sunset counterterms both scale like
\begin{equation}
\left( \Delta I\right)' = \frac{\Delta I}{\lambda v^2}\ .
\end{equation}
Note that we do set $k_B=c=1$ throughout.

We solve the Langevin equation on a cubic lattice 
using spatially periodic boundary conditions, 
lattice spacing $\ls$, and time discretization with time step $\lt$.
We shall compare results obtained using $\ls=0.4$ and $\ls=1$,
in dimensionless units. We always use $\lt=0.01$. On the
lattice, Eq.~(\ref{fluctuationdissipation}) is satisfied
by choosing $\xi$ independently at each discrete point
in space and time from a random distribution with 
$\langle\xi\rangle=0$ and $\langle\xi^2\rangle=2\eta(T)T/(\ls^3\lt)$.
We evolve the equations of motion using the fourth
order symplectic algorithm, although we have also
verified that using the second order 
algorithm suffices.
We have found that these algorithms introduce much less $\lt$-dependence
than the staggered leapfrog or stochastic Runge-Kutta algorithms.

\section{Order Parameter and Correlation Lengths}

\begin{figure}[t]
\begin{center}
\leavevmode
\psfig{file=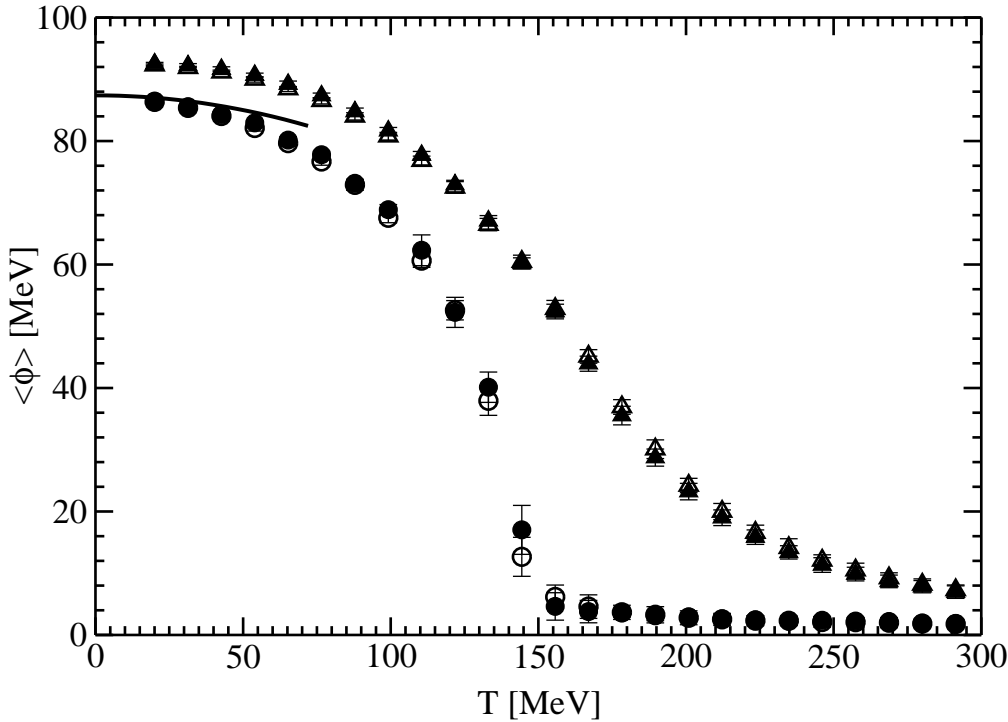,width=6in,angle=270,clip=,silent=}
\end{center}
\caption{The order parameter $\langle \phi \rangle$ as a
function of temperature $T$, for $H=0$ (circles)
and $H$ nonzero (triangles). Open and
filled symbols show results obtained with $\ls=1$ on a $26^3$ lattice
and $\ls=0.4$ on a $64^3$ lattice respectively, with counterterms
as described in the text.   
The line shows the prediction from chiral perturbation theory,
Eq. (\ref{chiral}). }
\label{phasediag}
\end{figure}
To analyze the transition
we have measured the order parameter $\langle \phi \rangle$ at several 
temperatures, as shown in Fig.~\ref{phasediag}. We define $\langle \phi \rangle$ by 
\begin{eqnarray}
\langle \phi \rangle = \sqrt{\sum_{a=1}^4 
\left( {1 \over N^3}\sum_{i,j,k=1}^N \phi_a(i,j,k)\right)^2}  ,
\end{eqnarray}
which is analogous to the magnetization in spin systems.
Error bars in the figure are the standard deviation 
from the mean in an ensemble of a few hundred measurements along
a thermalized Langevin trajectory. 
To determine after what time
the configuration has thermalized, we monitor the system
for equipartition 
and follow the order parameter until 
well after its behavior 
becomes non-monotonic.\footnote{To check for equipartition,
we only perform the simple test of verifying that the spatial average 
of all four $(\partial \phi_a/\partial t)^2$ are (like the order parameter) 
sufficiently  time-independent and are all given by $T$. This is equivalent to 
verifying that the kinetic energy of each field component is $T/2$.
In effect, we are measuring the temperature
of the fields, which we make sure agrees with the input value.}
For $H=0$ the system suffers from critical slowing down 
near its critical temperature,
and long Langevin trajectories are needed in order to obtain 
thermalization and a statistically meaningful set of measurements. 
For $H\neq 0$, thermalization occurs quickly at all temperatures, within
a time $t$ of  order $(10-100) \eta^{-1}$. 
Langevin algorithms may yield $a_t$-dependent 
results if $a_t$ is not
chosen small enough. 
To exclude this possibility, we checked our order parameter 
measurements against those obtained from configurations thermalized 
using a standard Metropolis Monte Carlo algorithm.

In Fig.~\ref{phasediag}, we have 
required that the behavior of $\langle \phi \rangle$ agrees
with the expectations from chiral perturbation theory at low 
temperatures~\cite{chpt},
\begin{equation}
\langle \phi \rangle = f_\pi \left( 1 - \frac{T^2}{12 f_\pi^2} \right)\ ,
\qquad T\ll f_\pi \ .
\label{chiral}
\end{equation}
In particular, we have enforced the absence of linear $T$-dependence
at small $T$. 
As anticipated in the previous section, 
this requires the addition of 
a further mass counterterm.
This counterterm is
not divergent in the $\ls\rightarrow 0$ limit, but it does
vary with $\ls$. It should be thought of as coming from
the order $a_s^p$ with $p\geq 0$  
contributions of infinitely many diagrams.     
We find that the linear $T$-dependence
at small $T$ is removed upon making the replacement
$-\lambda v^2 \rightarrow -\lambda v^2 + b_1 T$, with 
$b_1=0.425$ in dimensionless units.  Next, we have fixed the finite
counterterm proportional to $T^2$.  In principle, we could
have done this by enforcing agreement with the coefficient of the 
$T^2$ term in (\ref{chiral}). This is difficult in practice.
Instead, we note that in the absence of any finite
counterterm proportional to $T^2$ the
second order phase transition (for $H=0$) 
occurs at $T_c\simeq 130$ MeV, which is somewhat lower
than that expected in QCD~\cite{LatticeScreeningLengths,LatticeReviews}. 
We have pushed $T_c$
up to  $T_c\simeq 150$ MeV
by introducing 
$-\lambda v^2 \rightarrow -\lambda v^2 + b_1 T+ b_2 T^2$, 
with $b_2=-0.066$ in dimensionless 
units. These values of $b_1$ and $b_2$ were obtained with
a lattice spacing $\ls=0.4$ on a $64^3$ lattice, as shown
by the filled symbols in Fig.~2.  To
give the reader a sense of how they depend on $\ls$,
note that on a $26^3$ lattice with $\ls=1$ (and thus
the same physical volume as above) 
we find that with $b_2=0.084$ and $b_1$ unchanged from above, 
the order parameter as a function of temperature 
is the same as that for $\ls=0.4$ within error bars, as shown 
by the open symbols in Fig. 2.
We have verified that the $\ls$-dependence
of $b_2$ vanishes in the weak-coupling limit and is
non-divergent in the $\ls\rightarrow 0$ limit.

With $H=0$, the order
parameter shows a relatively sharp phase
transition at $T=T_c\simeq 150$~MeV,
which becomes sharper as the volume of the lattice is increased.
At nonzero $H$, the explicit symmetry breaking
favors the sigma direction $\langle\phi_a \rangle = \delta_{0a} f_\pi$ 
and the phase transition is smoothed into a crossover, which
occurs over a range of temperatures somewhat above the $T_c$
for $H=0$. 
We shall see below that this
crossover is centered at $T_{\rm cross}\simeq 180$~MeV.

\begin{figure}[t]
\begin{center}
\leavevmode
\psfig{file=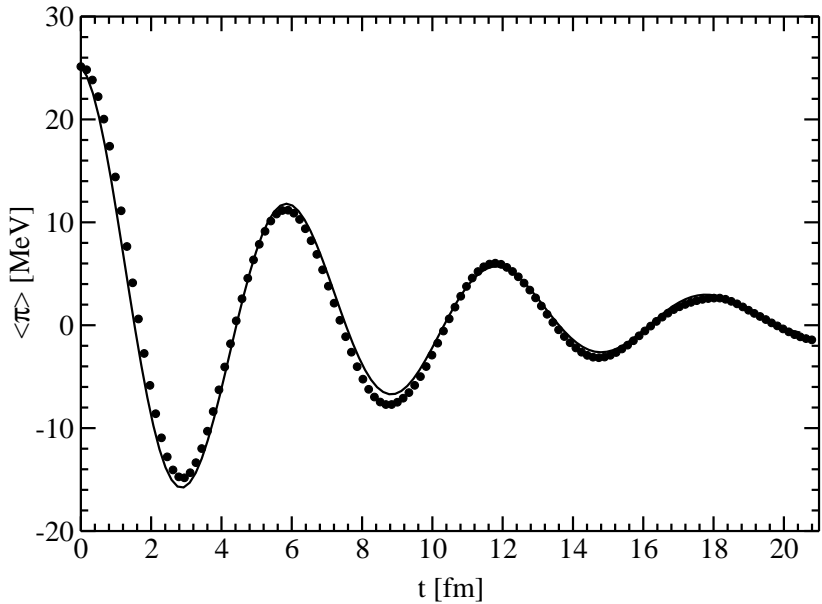,width=5in,angle=270,clip=,silent=}
\end{center}
\caption{An example of the fits used to determine $m_\pi$ and $m_\sigma$
as well as the associated decay rates. An instantaneous small perturbation
away from the thermal expectation values of the fields is induced  and 
its subsequent relaxation is fit to the form~(\ref{phifit}).  In
this example, we chose $\eta=0.05$
in dimensionless units.}
\label{massfit}
\end{figure}

\begin{figure}[t]
\begin{center}
\leavevmode
\psfig{file=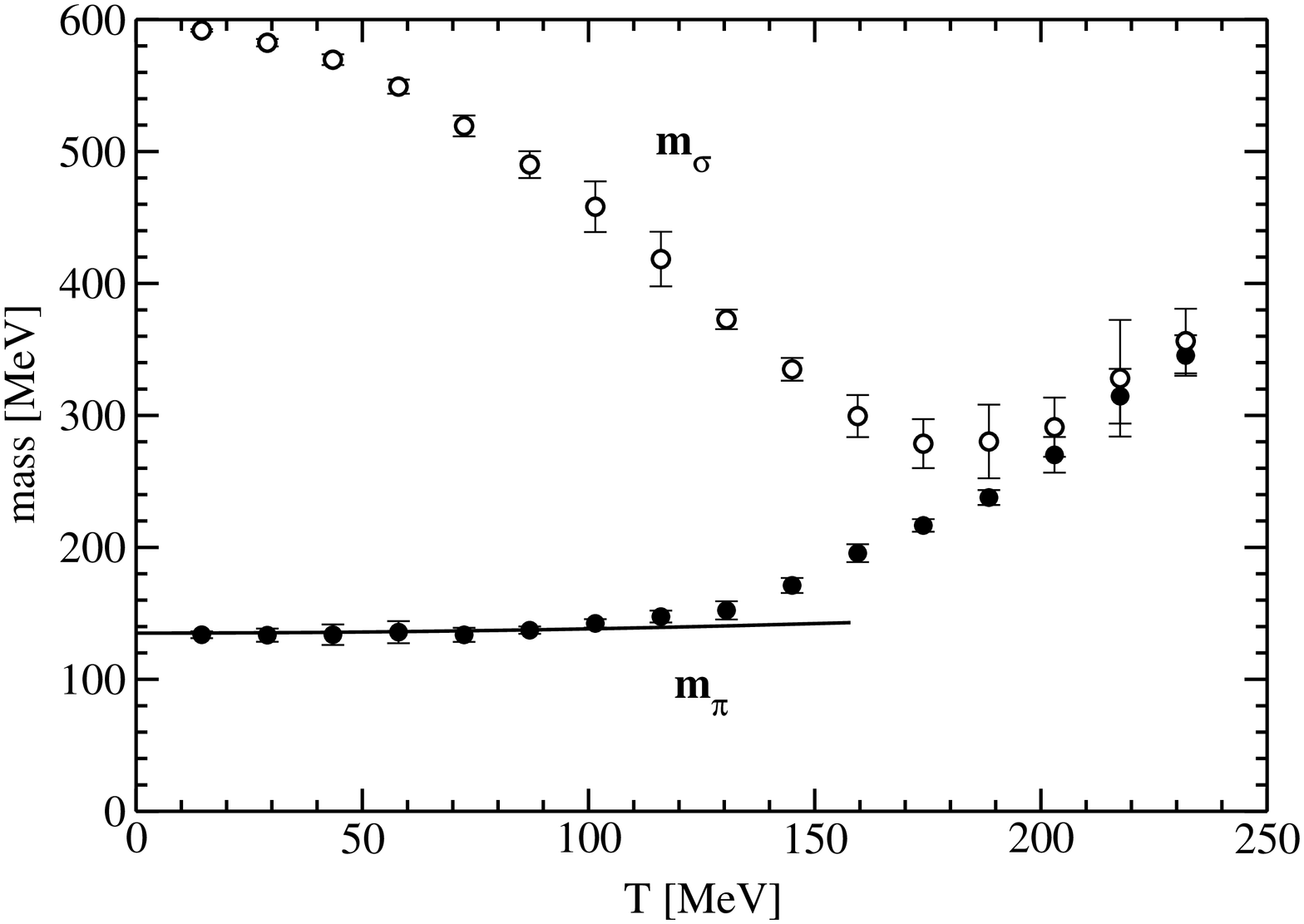,width=5in,angle=270,clip=,silent=}
\end{center}
\caption{The thermal masses of the $\sigma$ and $\pi$ fields 
as functions of temperature. Error bars represent statistical 
uncertainty in the ensemble of measurements and 
uncertainty in the fit, 
see Fig.\ref{massfit}. The line shows the prediction from 
chiral perturbation theory, Eq.~(\ref{mpichiralpt}).
The minimum of $m_\sigma$, which is a reasonable definition 
of the crossover temperature, is reached 
around $T_{\rm cross}\simeq  180$ MeV. All calculations were
done with $\ls=0.4$ on a $64^3$ lattice.}
\label{masses}
\end{figure}
In addition to measuring the order parameter we have also 
measured the temperature dependence of the masses of the pion and 
sigma fields.
To achieve this, we introduce a small amplitude displacement 
away from equilibrium in either the 
sigma or one of the pion field directions,
and time the resulting oscillations.
In practice, 
after first thermalizing the fields at a desired 
temperature as described above we 
add an offset in the value of the pion or 
sigma field to the thermalized field configuration, choosing the same
offset at every point in space. We typically shift
one of the pion fields by up to $0.1 f_\pi$ or shift the sigma field
by up to $0.05 f_\pi$. (A larger displacement of the sigma leads 
to non-linear response.)  We repeat this procedure
for ten different  
realizations of a canonical thermal ensemble. The time-dependent
response to the perturbation away from equilibrium is
most easily measured after first averaging 
$\langle\pi\rangle(t)$ or $\langle\sigma\rangle(t)$
over the ensemble of runs.
For small enough displacements we expect the response to be linear,
and characteristic of a damped harmonic oscillator. Thus
\begin{eqnarray}
\langle\phi_a\rangle 
= \langle\phi_a(t=0)\rangle \exp [- t/\tau] \cos(\omega t), \label{phifit}
\end{eqnarray}
with $\omega = \sqrt{m^2(T) - \tau^{-2}}$ for a spatially homogeneous 
perturbation. 
Taking $\eta$ small, in order to obtain
large relaxation times $\tau$,  
we have measured this behavior at 
many different temperatures. An example of the data and the fit to the 
form~(\ref{phifit}) is shown in Fig.~\ref{massfit}.
The results we obtain by following
this procedure at many temperatures
are shown in Fig.~\ref{masses}, 
together with the chiral perturbation
theory prediction for the pion mass~\cite{chpt}
\begin{equation}\label{mpichiralpt}
m_\pi^2(T)=m_\pi^2(0)\left(1+{T^2 \over 24 f_\pi^2}\right)\ ,
\qquad T\ll f_\pi \ .
\end{equation}
We thus confirm that with our
choice of counterterms, we are doing a good job of reproducing
the predictions of chiral perturbation theory at low temperatures.
We see that the sigma becomes lighter in the medium as a result of 
symmetry restoration at high temperatures. $m_\sigma$ displays a minimum
around $T_{\rm cross}\simeq 180$ MeV, which can be taken to be the 
crossover temperature. 
At $T_{\rm cross}$, $m_\pi\simeq 220$ MeV and $m_\sigma\simeq 280$ MeV.
At higher temperatures, $m_\sigma$ grows and
$m_\pi$ continues to grow monotonically. The two
masses are equal within error bars for 
$T\gtrsim 220$ MeV.

If nature were closer to the chiral
limit, $m_\pi$ and $m_\sigma$ would be less than $T$
for a range of temperatures near $T_{\rm cross}$,
justifying our use of classical field theory for
dynamics below.  As it is,
we shall be applying these methods somewhat beyond their
regime of quantitative validity.

\section{Disoriented Chiral Condensate Decay}

We have now characterized all the parameters and counterterms
in the model except $\eta(T)$.  $\eta$ plays no role 
in the equilibrium physics, but it is crucial to the dynamics.
In near-equilibrium dynamics, $\eta$ is the parameter
which controls the rate at which equilibrium is approached.
To complete the characterization of our linear sigma model, 
we investigate the linear response timescale $\tau$
for the decay of a long wavelength 
pion perturbation, excited as described in the previous
section.  For each temperature $T$, we extract $\tau$
by fitting the DCC decay profile to the form (\ref{phifit}).  
The $\tau$ so obtained depends on our choice of 
input $\eta$.\footnote{Note that we only measure $\tau$
for the decay of a pion perturbation. It may also be interesting
to investigate the timescale
$\tau_\sigma$ for the relaxation of a perturbation
in the sigma direction.  $\tau$ and $\tau_\sigma$ will 
differ at low temperatures and become degenerate above 
$T_{\rm cross}$.  In such an investigation, 
one could introduce 
$\eta_{\sigma\sigma}\neq\eta_{\pi\pi}$, and study
how $\tau$ and $\tau_\sigma$ depend on
$\eta_{\sigma\sigma}$ and $\eta_{\pi\pi}$.
Note that even though in this paper we take 
$\eta_{\sigma\sigma}=\eta_{\pi\pi}=\eta$,
this does not mean that $\tau$ and $\tau_\sigma$ are equal.}

The timescale $\tau(T)$ for the decay of 
a region of disoriented chiral condensate in
contact with an equilibrated gas of hadrons has
been computed previously by Steele and Koch~\cite{SteeleKoch}.
We shall seek to choose $\eta(T)$ so that the $\tau(T)$ 
we then extract reproduces that calculated as 
in Ref.~\cite{SteeleKoch}.
In so doing, we are normalizing the magnitude of the 
dissipation in the Langevin evolution to a physical observable.

The dissipation parameter $\eta(T)$ has been  computed perturbatively by 
Rischke~\cite{Rischke:1998qy} from the imaginary part of the two-loop 
pion self-energy in the finite temperature quantum linear sigma model, with
the result:
\begin{eqnarray}\label{Rischkeeta}
\eta(T)&=& \left(\frac{4\lambda f_\pi}{N} \right)^2 
\frac{m_\sigma^2}{4\pi m_\pi^3} 
\sqrt{1-\frac{4m_\pi^2}{m_\sigma^2}}
\frac{1-e^{-m_\pi/T}}{1-e^{-m_\sigma^2/2m_\pi T}}
\frac1{e^{(m_\sigma^2-2m_\pi^2)/2m_\pi T}-1}.
\end{eqnarray}
At $\lambda=20$, we can only expect this calculation to
be a qualitative guide.

\begin{figure}[t]
\begin{center}
\leavevmode
\psfig{file=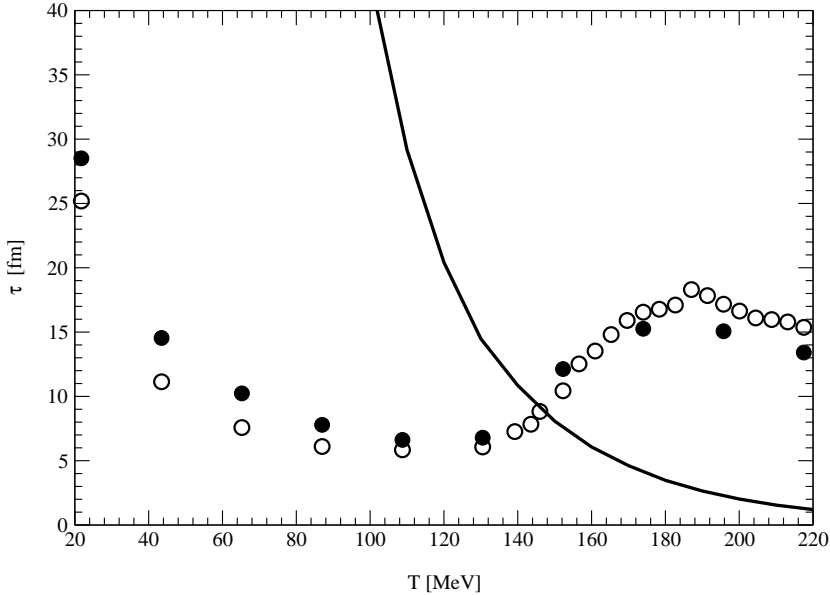,width=5in,angle=270,clip=,silent=}
\end{center}
\caption{The solid curve shows the 
decay time of a spatially homogeneous DCC
via interaction with a gas of hadrons, computed
following Ref.~\protect\cite{SteeleKoch} as described in the 
text.  The open (filled)
circles show $\tau_{\rm cl}(T)$, obtained
from the Langevin equation with $\eta=0$, for $\ls=1$ ($\ls=0.4$).  
The uncertainty in each point (statistical and that from the
fit) is comparable to the dispersion between open and
closed circles.
$\tau_{\rm cl}(T)$ reflects the dissipation of the long
wavelength DCC induced by
contact with thermalized classical fields at shorter wavelengths. 
We see that $\tau_{\rm cl} (T) < \tau(T)$ below $T\simeq 145$~MeV.}
\label{taus_lowT}
\end{figure}

We shall slightly extend the calculation of $\tau(T)$ performed
by Steele and Koch, and so instead of simply quoting
the result we first sketch their computation. 
We wish to analyze a large, spatially uniform, DCC
in contact with a thermal gas of many other hadronic
species  $X=\pi,\,K,\,...$,   
which will interact with the DCC and ultimately cause it to decay.  
Modeling the DCC by a coherent state, and working through the 
LSZ formalism, Steele and Koch show that the depletion of the particles in the 
DCC, $dN/dt$, is proportional to the square of the interaction 
amplitudes~\cite{SteeleKoch}:
\begin{eqnarray}
\frac{dN}{dt} = \int\! {d^3 k_1 \over (2 \pi)^3} \, 
{d^3  k_2 \over (2 \pi)^3}\, {d^3 k_3 \over (2 \pi)^3}\, 
\sum_X
F^X_{123}\,  
\langle | {\cal T}_{\pi X} |^2\rangle 
2\pi\, \delta (E_0+E_1-E_2-E_3) \,
\frac{|\zeta_{k_2+k_3-k_1}|^2}{(2E_0)^2}  \, .
\label{prerate}
\end{eqnarray}
For $\pi$-$\pi$ scattering, the thermal weighting is given by 
\begin{equation}
F^\pi_{123}=f_2 f_3(1+f_1)-f_1 (1+f_2)(1+f_3) ,
\end{equation}
with $f_i=(\exp(E_i/T)-1)^{-1}$ representing the Bose-Einstein
momentum distribution.  
The DCC pion energy is $E_0=[( k_2+ k_3- k_1)^2+m_\pi^2]^{1/2}$.
$|\zeta|^2$ to first approximation gives the momentum-conserving
delta-function leading to a formula for the half-life $\tau$ of 
the DCC 
\begin{equation}
\!\!\!\frac1{N} \frac{dN}{dt}\!\! =\! \frac1{2E_0}\! 
\int\!\!  {d^3 k_1 \over (2 \pi)^3} \, 
{d^3  k_2 \over (2 \pi)^3}\, {d^3 k_3 \over (2 \pi)^3}\,
\sum_X
F^X_{123}\, \langle |{\cal T}_{\pi X}|^2\rangle \,
(2\pi)^4 \delta^4(\Sigma\, k_i) 
\, \equiv - \frac1{\tau}
\label{rate}
\end{equation}
The focus of Ref.~\cite{SteeleKoch} was on the late time decay
of a DCC, due to a heat bath with a temperature well below $T_c$.
In this context it is reasonable to 
take the scattering amplitudes ${\cal T}_{\pi X}$ 
from phase-shifts of real zero-temperature data. 
If these are not known, then a 
Breit-Wigner form can be fit to the dominant resonance.
To obtain the values of $\tau(T)$ shown below, 
$\pi\pi$ was taken 
from data, $\pi K$ was modeled by the $K^*(892)$ resonance, $\pi\rho$ 
was modeled by the $a_1$ resonance, $\pi N$ was 
taken from data, 
$\pi \bar N$ was taken from the same 
$\pi N$ data, and $\pi \Delta$ ($\pi\bar\Delta$)
was modeled by the $N^*(1675)$ resonance.  
Note that at $T=170$~MeV the pions account for only
about half the particles in our equilibrated resonance
gas. All the other species together do play an
important role.
We have extended the previous analysis by
including the   
dissipation due to $\pi K$, $\pi \Delta$ and $\pi \bar N$
scattering, not included in Ref.~\cite{SteeleKoch}.
The $\tau(T)$  obtained in this way is shown in
Fig. \ref{taus_lowT}.  The calculation is only quantitatively
valid below
$T_c$, but we have plotted the results up to higher temperatures
also.

\begin{figure}[t]
\begin{center}
\leavevmode
\psfig{file=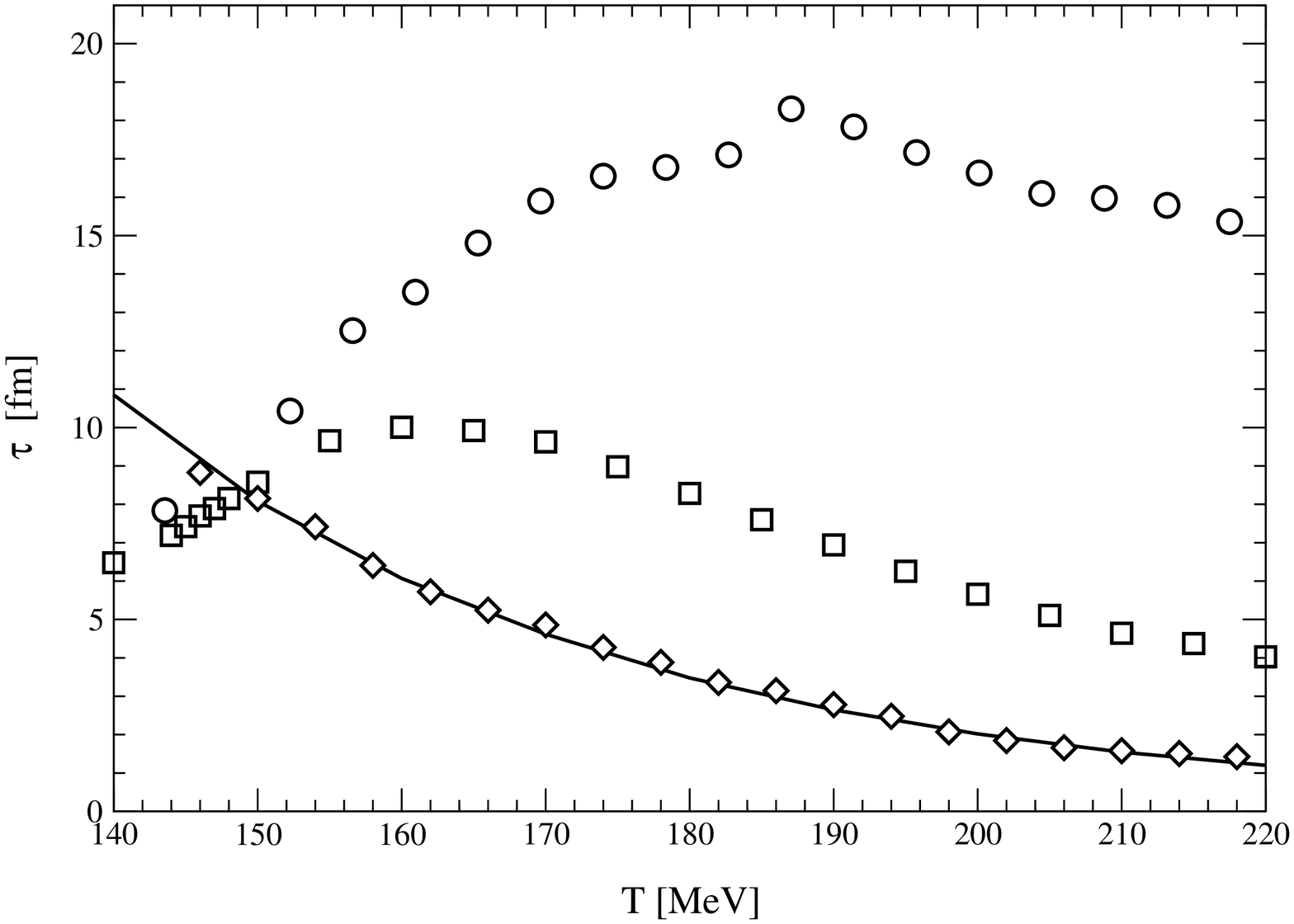,width=5in,angle=270,clip=,silent=}
\end{center}
\caption{Circles show the DCC decay time $\tau_{\rm cl}(T)$ 
for the  classical $\sigma$-model at 
$\eta=0$.   Squares show $\tau(T)$ obtained
from the Langevin equation with $\eta(T)$ 
taken from Rischke's perturbative calculation.
Diamonds show $\tau(T)$ obtained
from the Langevin equation with $\eta(T)$ chosen
so as to reproduce $\tau(T)$ from the calculation
of Steele and Koch, shown as the solid line. Errors
are similar to those discussed for Fig.~5.}
\label{taus}
\end{figure}

Below we shall choose $\eta(T)$ to match the decay time $\tau(T)$ of a 
spatially homogeneous DCC
in our Langevin evolution to the values calculated by Steele and Koch. 
First, however, let us see what we obtain with $\eta=0$, as
in Fig. \ref{taus_lowT}.  That is, we first thermalize
at a given temperature, add a DCC, and then
set $\eta=0$ thus removing
the classical
fields from further contact with any heat bath.  In so doing, however,
we do not remove all sources of dissipation.  The long 
wavelength DCC still decays, by virtue of its coupling
to those finite wavelength modes which we are
treating classically.  That is,  the DCC decays via contact with an 
effective classical heat bath, present even for $\eta=0$.  We see that at low
temperatures the resulting $\tau_{\rm cl}(T)$ 
turns out to be smaller
than the desired $\tau(T)$, namely that due to 
contact with a gas of hadrons.  The reason is that $\tau(T)$
becomes very large at low temperatures, due to the exponential
suppression of the thermal occupation numbers in the (quantum)
gas of hadrons.  The classical thermal field theory overestimates the
occupation numbers, and thus the dissipation, leading to an underestimate 
of $\tau$.

Note that we have checked that
our calculation of $\tau_{\rm cl}$ is independent of the lattice
spacing $a_s$. As shown in Fig.~5, we have compared 
results obtained on a $64^3$ 
lattice with $\ls=0.4$ to those obtained on a $26^3$ lattice with $\ls=1$.
As described
in the previous section, the ultraviolet-finite
counterterms must be chosen differently in these two
cases in order that the equilibrium behavior of the order
parameter is the same in both cases.
With
the counterterms so adjusted, $\tau_{\rm cl}(T)$ changes
little with $\ls$.

The shape of $\tau_{\rm cl}(T)$, shown in Fig.~\ref{taus_lowT}, 
is interesting and may be understood as follows.  
$\tau_{\rm cl}(T)$ should grow like $1/T$ as $T\rightarrow 0$ (whereas
$\tau$ grows much more rapidly in this limit). 
We do find that 
$\tau_{\rm cl}$ rises at the lowest temperatures 
we have explored. The minimum of   $\tau_{\rm cl}$
coincides with the minimum value of the ratio  $m_\pi(T)/T$, which
occurs at $T\simeq 120-130$~MeV.  This means that the classical
thermal pions are most numerous in this temperature regime, making
it plausible that the classical dissipation
is largest and $\tau_{\rm cl}$ is smallest.

We now see the challenge posed by our use of a classical theory,
with its intrinsic decay time $\tau_{\rm cl}(T)$.  By turning
on a nonzero $\eta$, we can only add dissipation.  This means
that we can only reduce $\tau$, relative
to $\tau_{\rm cl}(T)$.  At temperatures where
$\tau_{\rm cl}(T)< \tau(T)$,
therefore, there is no way to use our Langevin
equation to reproduce the correct dynamics of the decay 
of a DCC.  Thus, we find that our model can only
be used to describe the long wavelength dynamics
in the presence of a heat bath with  $T \gtrsim 145$ MeV,
corresponding to about 80\% of the crossover temperature 
$T_{\rm cross}$.  $T \sim 145$ MeV is
large enough that the assumptions in
the calculation of $\tau$ as in 
Ref.~\cite{SteeleKoch} may be starting to break
down, which means that
our estimate of the limit of validity of our
analysis may have a little play in it. However, the fundamental
fact that $\tau$ should rise rapidly with decreasing
temperature while $\tau_{\rm cl}$ does not is incontrovertible.

\begin{figure}[t]
\begin{center}
\leavevmode
\psfig{file=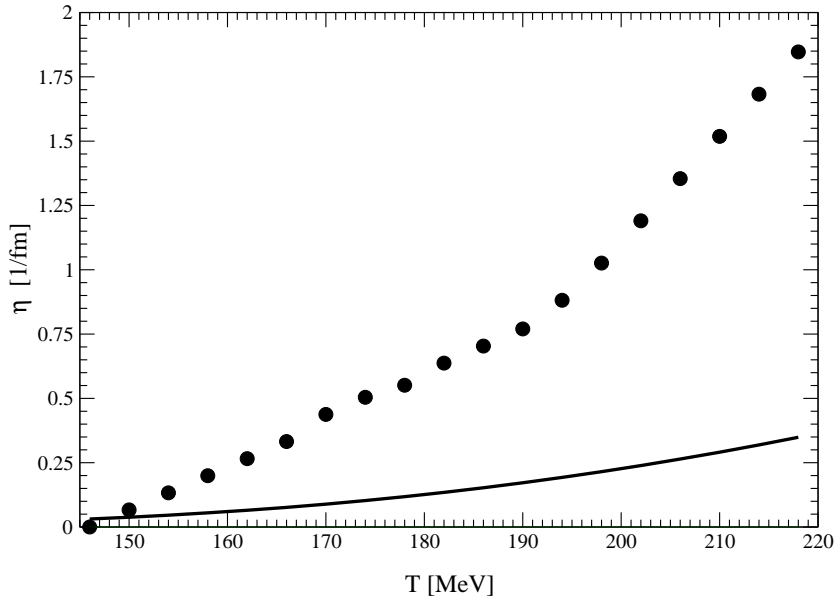,width=5in,angle=270,clip=,silent=}
\end{center}
\caption{$\eta(T)$ from Rischke's perturbative calculation (solid line)
and $\eta(T)$  chosen so that 
the Langevin equation
reproduces $\tau(T)$ computed by Steele and Koch (circles).}
\label{etas}
\end{figure}
Above $T\simeq 145$~MeV, we can proceed as planned.  We
can choose an $\eta(T)$ in such a way as to reproduce
the predictions of Steele and Koch, as demonstrated 
in Fig.~\ref{taus}.  The required $\eta(T)$
is plotted in Fig.~\ref{etas}, where it is also compared
to the $\eta(T)$ of Eq. (\ref{Rischkeeta}), calculated perturbatively by 
Rischke~\cite{Rischke:1998qy}. 
We see (from both Figs. \ref{taus} and \ref{etas})
that introducing the perturbatively calculated $\eta(T)$ into
the Langevin equation is not sufficient to obtain
a DCC decay timescale $\tau(T)$ in agreement with that
induced via contact with a thermal hadron 
gas.\footnote{It is not clear whether we should
have expected the use of $\eta$ from (\ref{Rischkeeta})
to yield too rapid or too slow dissipation.  On the one
hand, 
we may have expected that use of the  perturbative $\eta(T)$
in the Langevin equation would yield excess dissipation,
as there is always classical dissipation in addition. On
the other hand,
we may have expected that it
would introduce insufficient dissipation, as its calculation
neglects the effects of all hadrons in the heat bath except the 
pions and sigma.}   
Both the calculations of Rischke and of Steele and Koch
are only of quantitative validity below $T_c$, as they
use zero temperature masses and interactions.   Nevertheless,
the discrepancy between the $\eta(T)$'s in the vicinity of 
$T_c$ is sufficiently
large that we take this to mean that, for
$\lambda=20$, the perturbative 
calculation underpredicts $\eta$.

\section{Discussion}

We have shown how to use
a stochastic classical linear sigma model
to obtain a reasonable description
of the equilibrium thermodynamics and the long-wavelength 
near-equilibrium dynamics of QCD
at temperatures of order that of its phase transition. 
The use of any classical statistical field theory
around the chiral transition cannot be quantitatively
justified, because there are no excitations with masses
which are very small compared to the temperature.
Nevertheless, we have determined the choice of divergent
counterterms needed to remove the ultraviolet divergences
from the classical linear sigma model, making the
long distance physics described therein 
independent of the lattice spacing.  We have then been
able to find the finite counterterms needed in order
for the equilibrium temperature dependence of the order parameter
in the linear sigma model to be as in QCD
at low temperatures, as calculated in chiral
perturbation theory, and for $T_c$ to be as in QCD,
as calculated in lattice gauge theory.
The temperature dependence of the pion and sigma masses then
come out nicely in accord with expectations. 
We have then fixed $\eta(T)$ such that the timescale for the
dissipation of a DCC in our Langevin model
agrees with that calculated previously.
The main constraint on our model is that it cannot
be used at low temperatures, where the classical dissipation
is too great even when $\eta$ is set to zero.   We find
that this limits its use to temperatures above about
80\% that of the crossover transition itself.
Above this temperature, we show how to choose $\eta(T)$ in
such a way that 
the long wavelength
dynamics of the pions and sigma should be 
reproduced reasonably well in our model, 
at least as long as the system 
is not too far out of thermal equilibrium. 

In this work, we have used the dynamics of an idealized
DCC as a device with which to normalize our model.
In subsequent work, we shall use the now normalized
model to investigate the
dynamical setting where the system 
expands and cools through
its chiral crossover transition.
We hope to determine with some confidence how fast the expansion
and cooling must be in order for the dynamics to
be driven significantly from equilibrium.
We will then follow the
nonequilibrium long wavelength evolution for as long
as occupation numbers remain reasonably high.
Unlike in the idealized setting utilized for convenience in this paper 
and in the analytic computations of $\eta(T)$, the long wavelength 
modes will certainly not describe a spatially uniform 
condensate.

\section*{Acknowledgements}

KR is grateful to D. Rischke for long discussions several years ago
which started DR down the path toward Ref.~\cite{Rischke:1998qy}
and started KR down the path toward the present paper.
We also acknowledge conversations with A. K. Chaudhuri.
Numerical work was performed in part at the T-Division/CNLS Avalon cluster at 
Los Alamos National Laboratory. This work is supported in part 
by the Department of Energy under 
cooperative research agreement \#DF-FC02-94ER40818.
The work of KR was supported in part by a DOE OJI Award and by the
A. P. Sloan Foundation.


\begin{thebibliography}{99}

\bibitem{MuellerSon} R.~Baier, A.~H.~Mueller, D.~Schiff, and D.~T.~Son,
Phys.\ Lett.\ B {\bf 502}, 51 (2001) [hep-ph/0009237].

\bibitem{EllipticFlow}
K.~H.~Ackermann {\it et al.}  [STAR Collaboration],
Phys.\ Rev.\ Lett.\  {\bf 86}, 402 (2001)
[nucl-ex/0009011].


\bibitem{earlycritptpapers}
A. Barducci, 
R. Casalbuoni, S. DeCurtis, R. Gatto, G. Pettini, Phys. Lett. B {\bf 231},
463 (1989);
S.P. Klevansky, Rev. Mod. Phys. {\bf 64}, 649 (1992); 
A. Barducci, R.~Casalbuoni, G. Pettini and R. Gatto, Phys. Rev. D {\bf 49},
426 (1994);
M. Stephanov, Phys. Rev. Lett. {\bf 76}, 4472 (1996);
Nucl. Phys. Proc. Suppl. {\bf 53}, 469 (1997);
J.~Berges and K.~Rajagopal,
Nucl.\ Phys.\ B {\bf 538}, 215 (1999)
[hep-ph/9804233];
M.~A.~Halasz, A.~D.~Jackson, R.~E.~Shrock, 
M.~A.~Stephanov and J.~J.~Verbaarschot,
Phys.\ Rev.\ D {\bf 58}, 096007 (1998)
[hep-ph/9804290].

\bibitem{SRS1}
M.~Stephanov, K.~Rajagopal and E.~Shuryak,
Phys.\ Rev.\ Lett.\  {\bf 81}, 4816 (1998)
[hep-ph/9806219].

\bibitem{SRS2}
M.~Stephanov, K.~Rajagopal and E.~Shuryak,
Phys.\ Rev.\ D {\bf 60}, 114028 (1999)
[hep-ph/9903292].

\bibitem{critptreview}
For a review see K.~Rajagopal,
Acta Phys.\ Polon.\ B {\bf 31}, 3021 (2000)
[hep-ph/0009058].



\bibitem{Bubbling}
K.~Rajagopal,
Nucl.\ Phys.\ A {\bf 680}, 211 (2000)
[hep-ph/0005101].

\bibitem{HeiselbergJacksonMishustin}
See, for example,
H.~Heiselberg and A.~D.~Jackson,
nucl-th/9809013;
I.~N.~Mishustin,
Phys.\ Rev.\ Lett.\  {\bf 82}, 4779 (1999)
[hep-ph/9811307].


\bibitem{BerdnikovRajagopal}
B.~Berdnikov and K.~Rajagopal,
Phys.\ Rev.\ D {\bf 61}, 105017 (2000)
[hep-ph/9912274].




\bibitem{RajagopalWilczekDCC}
K.~Rajagopal and F.~Wilczek,
Nucl.\ Phys.\  B {\bf 404}, 577 (1993)
[hep-ph/9303281].

\bibitem{AnselmRyskin}
A.~A.~Anselm, Phys. Lett. B {\bf 217}, 169 (1988);
A.~A.~Anselm and M.~G.~Ryskin,
Phys.\ Lett.\  B {\bf 266}, 482 (1991).

\bibitem{BlaizotKrzywicki}
J.~Blaizot and A.~Krzywicki,
Phys.\ Rev.\ D {\bf 46}, 246 (1992).

\bibitem{Bjorken}
J. D. Bjorken, Int. J. Mod. Phys. {\bf A7}, 246 (1992);
{\it ibid.}, 4189;
J.~D.~Bjorken,
Acta Phys.\ Polon.\ B {\bf 23}, 637 (1992);
K.~L.~Kowalski and C.~C.~Taylor,
hep-ph/9211282;
J.~D.~Bjorken, K.~L.~Kowalski and C.~C.~Taylor,
hep-ph/9309235.

\bibitem{RajagopalWilczekStatic}
K.~Rajagopal and F.~Wilczek,
Nucl.\ Phys.\  B {\bf 399}, 395 (1993)
[hep-ph/9210253].

\bibitem{WA98}
M.~M.~Aggarwal {\it et al.}  [WA98 Collaboration],
Phys.\ Lett.\ B {\bf 420}, 169 (1998)
[hep-ex/9710015].
M.~M.~Aggarwal {\it et al.}  [WA98 Collaboration],
Phys.\ Rev.\ C {\bf 64}, 011901 (2000)
[nucl-ex/0012004].

\bibitem{OtherSignatures}
C.~Greiner, C.~Gong and B.~Muller,
Phys.\ Lett.\ B {\bf 316}, 226 (1993)
[hep-ph/9307336].
Z.~Huang, M.~Suzuki and X.~Wang,
Phys.\ Rev.\ D {\bf 50}, 2277 (1994)
[hep-ph/9403300];
S.~Gavin, Nucl. Phys. A {\bf 590}, 163 (1995);
hep-ph/9407368;
Z.~Huang, I.~Sarcevic, R.~Thews and X.~Wang,
Phys.\ Rev.\ D {\bf 54}, 750 (1996)
[hep-ph/9511387];
Z.~Huang and X.~Wang,
Phys.\ Lett.\ B {\bf 383}, 457 (1996)
[hep-ph/9604300];
D.~Boyanovsky, H.~J.~de Vega, R.~Holman and S.~Prem Kumar,
Phys.\ Rev.\ D {\bf 56}, 5233 (1997)
[hep-ph/9701360];
Phys.\ Rev.\ D {\bf 56}, 3929 (1997)
[hep-ph/9703422];
Y.~Kluger, V.~Koch, J.~Randrup and X.~Wang,
Phys.\ Rev.\ C {\bf 57}, 280 (1998)
[nucl-th/9704018];
H.~Hiro-Oka and H.~Minakata,
Phys.\ Lett.\ B {\bf 425}, 129 (1998)
[Erratum-ibid.\ B {\bf 434}, 461 (1998)]
[hep-ph/9712476];
T.~C.~Petersen and J.~Randrup,
Phys.\ Rev.\ C {\bf 61}, 024906 (2000)
[nucl-th/9907051];
C.~Chow and T.~D.~Cohen,
Phys.\ Rev.\ C {\bf 60}, 054902 (1999)
[nucl-th/9908013];
J.~I.~Kapusta and S.~M.~Wong,
Phys.\ Rev.\ Lett.\  {\bf 86}, 4251 (2001)
[nucl-th/0012006].

\bibitem{NonEqbmQFT} 
F.~Cooper, S.~Habib, Y.~Kluger, E.~Mottola, J.~P.~Paz and P.~R.~Anderson,
Phys.\ Rev.\ D {\bf 50}, 2848 (1994)
[hep-ph/9405352];
D.~Boyanovsky, H.~J. de Vega, and R.~Holman, 
Phys.\ Rev.\ D {\bf 51}, 734 (1995) [hep-ph/9401308]; 
F. Cooper, Y.~Kluger, E. Mottola,
and J. Paz, Phys.\ Rev.\ D {\bf 51}, 2377 (1995) [hep-ph/9401308]; 
M.~A.~Lampert, J.~F.~Dawson and F.~Cooper,
Phys.\ Rev.\ D {\bf 54}, 2213 (1996)
[hep-th/9603068];
F. Cooper, Y. Kluger, E. Mottola, Phys.\ Rev.\ C {\bf 54}, 3298 (1996) 
[hep-ph/9604284];
D.~Boyanovsky, F.~Cooper, H.~J.~de Vega and P.~Sodano,
Phys.\ Rev.\ D {\bf 58}, 025007 (1998)
[hep-ph/9802277].


\bibitem{AHW}
M.~Asakawa, Z.~Huang and X.~Wang,
Phys.\ Rev.\ Lett.\  {\bf 74}, 3126 (1995)
[hep-ph/9408299].

\bibitem{SteeleKoch} J.V.~Steele, and  V.~Koch, 
Phys.~Rev.~Lett. 
{\bf 81}, 4096 (1998) [nucl-th/9806055]. 

\bibitem{Rischke:1998qy}
D.~H.~Rischke,
Phys.\ Rev.\  C {\bf 58}, 2331 (1998)
[nucl-th/9806045].

\bibitem{Greiner:1997dx}
C.~Greiner and B.~Muller,
Phys.\ Rev.\ D {\bf 55}, 1026 (1997)
[hep-th/9605048].


\bibitem{XuGreiner} Z.~Xu and C.~Greiner, 
Phys.\ Rev.\ D {\bf 62}  036012 (2000) [hep-ph/9910562].

\bibitem{BiroGreiner}
T.~S.~Biro and C.~Greiner,
Phys.\ Rev.\ Lett.\  {\bf 79}, 3138 (1997)
[hep-ph/9704250].

\bibitem{Chaudhuri1} A.~K.~Chaudhuri, Phys.\ Rev.\ D {\bf 59},
117503 (1999) [nucl-th/9809018]; 
J. Phys. G {\bf 27} 175 (2001) [hep-ph/9908376]; hep-ph/0007332; 
hep-ph/0011003.

\bibitem{Chaudhuri2} A.~K.~Chaudhuri, hep-ph/0105013.

\bibitem{Randrup} J.~Randrup, 
Phys.\ Rev.\ Lett.\ {\bf 77}, 1226 (1996) [hep-ph/9605223];
Nucl.\ Phys.\  A {\bf 616}, 531 (1997) [hep-ph/9612453];
J.~Randrup,
Phys.\ Rev.\ C {\bf 62}, 064905 (2000)
[nucl-th/0010104].




\bibitem{PisarskiWilczek}
R. Pisarski and F. Wilczek, Phys. Rev. D {\bf 29}, 338 (1984);
F. Wilczek, Int. J. Mod. Phys. A {\bf 7}, 3911 (1992).


\bibitem{LatticeScreeningLengths}
S. Gottlieb et al., 
Phys. Rev. D {\bf 55}, 6852 (1997) [hep-lat/9612020].



\bibitem{Pisarski}
R.~D.~Pisarski,
Phys.\ Rev.\ D {\bf 62}, 111501 (2000)
[hep-ph/0006205];
A.~Dumitru and R.~D.~Pisarski,
Phys.\ Lett.\ B {\bf 504}, 282 (2001)
[hep-ph/0010083].


\bibitem{Gupta}
S.~Datta and S.~Gupta,
Nucl.\ Phys.\ B {\bf 534}, 392 (1998) [hep-lat/9806034].


\bibitem{GavaiGupta}
R. V. Gavai and S. Gupta, Phys. Rev. Lett. {\bf 85}, 2068 (2000)
[hep-lat/0004011].

\bibitem{LatticeReviews}
For reviews, see 
F.~Karsch,
Nucl.\ Phys.\ Proc.\ Suppl.\  {\bf 83}, 14 (2000)
[hep-lat/9909006];
E. Laermann,  Nucl. Phys. Proc. Suppl. {\bf 63}, 114 (1998);
and A. Ukawa, Nucl. Phys. Proc. Suppl. {\bf 53}, 106 
(1997).


\bibitem{Bett} L.~M.~A.~Bettencourt, 
Phys.\ Rev.\ D {\bf 63}, 045020 (2001) [hep-ph/0005264].

\bibitem{HoHa} P.C.~Hohenberg and B.I.~Halperin, 
Rev.\ Mod.\ Phys. {\bf 49}, 435 (1977).


\bibitem{Parisi} G.~Parisi, {Statistical Field Theory}
(Addison-Wesley, New  York, 1988). 

\bibitem{Farakos} K. Farakos, K. Kajantie, K. Rummukainen, M. Shaposhnikov,
Nucl.\ Phys. B {\bf 425}, 67 (1994) [hep-ph/9404201].

\bibitem{chpt}
P.~Gerber and H.~Leutwyler, Nucl.\ Phys.\ B {\bf  321}, 387 (1989);
R.~D.~Pisarski and M.~Tytgat,
Phys.\ Rev.\ D {\bf 54}, 2989 (1996)
[hep-ph/9604404].



\end{thebibliography}
\end{document}